\documentstyle[12pt,epsfig]{article}
\newcommand{\be}{\begin{equation}}
\newcommand{\ee}{\end{equation}}
\newcommand{\bear}{\begin{eqnarray}}
\newcommand{\eear}{\end{eqnarray}}
\newcommand{\ba}{\begin{array}}
\newcommand{\ea}{\end{array}}

\newskip\humongous \humongous=0pt plus 1000pt minus 1000pt

\newif\ifdtup

\def\oldreffmt#1{\rlap{[#1]} \hbox to 2\parindent{}}

\def\figfmt#1{\rlap{Figure {#1}} \hbox to 1in{}}

\def\etal{\hbox{\it et al.}}

\def\slash#1{#1\!\!\!/\!\,\,}
\def\beq{\begin{equation}}
\def\eeq{\end{equation}}
\def\bea{\begin{eqnarray}}
\def\eea{\end{eqnarray}}
\def\half{\frac{1}{2}}

\def\bq{\begin{quote}}
\def\eq{\end{quote}}

\def\half{\frac{1}{2}} 
\def \lta {\mathrel{\vcenter
     {\hbox{$<$}\nointerlineskip\hbox{$\sim$}}}}
\def \gta {\mathrel{\vcenter
     {\hbox{$>$}\nointerlineskip\hbox{$\sim$}}}} 

\def \etal {{\it et al.}\ }
\relax

\newdimen\tdim
\tdim=\unitlength
\def\bar{\overline}

\begin{document}

\pagestyle{empty}
\begin{titlepage}
\def\thepage {}    
         \title{   \vspace*{1.5cm}  \bf  
Natural Theories of Ultra--Low Mass PNGB's:
Axions and Quintessence } 
\author{  \\ [0.5cm] 
\bf  Christopher T. Hill \\[2mm]
\bf Adam K. Leibovich \\ [2mm]
{\small {\it Fermi National Accelerator Laboratory}}\\  
{\small {\it P.O. Box 500, Batavia, Illinois 60510, USA}}  
\thanks{  
hill@fnal.gov, adam@fnal.gov }\\  
}  
\date{\today}  
\maketitle  
\vspace*{-10.75cm}  
\noindent  
\begin{flushright}
FERMILAB-Pub-02/085-T \\ [1mm]
May, 2002
\end{flushright}

\vspace*{11.1cm}
\baselineskip=18pt

\begin{abstract}

  {\normalsize
We consider the Wilson Line PNGB
which arises in a $U(1)^N$ gauge theory, abstracted
from a latticized, periodically
compactified extra dimension $U(1)$.
Planck scale breaking of the PNGB's
global symmetry is suppressed,  
providing natural candidates for the axion
and quintessence. We construct an
explicit model in which the axion
may be viewed as the $5$th component
of the $U(1)_Y$
gauge field in a $1+4$ latticized
periodically compactified extra dimension.  
We also construct a quintessence PNGB model where the
ultra-low mass arises from
Planck-scale suppressed physics itself.
}
\end{abstract}

\vfill
\end{titlepage}

\baselineskip=18pt
\renewcommand{\arraystretch}{1.5}  
\pagestyle{plain}
\setcounter{page}{1}  
  
\section{Introduction} 

Ultra--low mass spin-$0$ fields
are desired in particle physics, and often carry
cosmological implications.
As usual,
``low-mass'' means a mass scale hierarchically
small compared to much
larger mass scales in the problem. The Planck mass,
$M_P$, is the largest mass scale present in any field theory
problem.  Achieving a ``low-mass''
spin-$0$ field is generally challenging, owing at
least to putative Planck scale quantum effects, and always requires
the judicious use of symmetry.
Key examples of these are the Higgs boson
and the axion.  We will not recite the
issues associated with electroweak symmetry
breaking and the Higgs boson naturalness
problem presently, though it has been the
main driver of speculation about physics beyond the weak scale
(e.g., see the recent reviews,
\cite{ellis,hillsimm}). Rather, we focus presently on
the axion, and ultra-low-mass pseudo-Nambu-Goldstone bosons (PNGB's).

The axion was proposed as a massless particle, in the absence of QCD  
effects, which arises as a Nambu-Goldstone-boson (NGB) when the $U(1)$  
global Peccei-Quinn symmetry (PQ-symmetry)
is spontaneously broken at a very high mass scale $f$ \cite{pq,axion}.
We also have small explicit symmetry breaking effects which
arise from QCD, via the axial vector current anomaly,
which are of order
$\sim \Lambda_{QCD}$.  The axion develops a residual
mass of order $m_{axion} \sim \Lambda^2_{QCD}/f$,
becoming a pseudo-Nambu-Goldstone-boson (PNGB).
With $f\sim 10^{12}$ GeV this is a tiny mass
scale of order $10^{-5}$ eV. At the minimum of its potential,
the VEV of the axion identically cancels the unwanted $\theta$
angle of QCD to a precision of $ \leq 10^{-9}$.

This picture can be criticized as follows. The PQ-symmetry
is postulated as a global symmetry at all scales
above $\Lambda_{QCD}$.  However, at the scale $M_P$ we expect
quantum gravity (string theory) to become important. The
vacuum at this scale will contain large fluctuations involving
the formation and rapid evaporation of tiny black-holes.
Black holes can eat global symmetries, i.e., fields carrying
global charges will experience global charge nonconserving
interactions in the presence of mini-black-holes. Hence we expect
large explicit PQ--symmetry breaking effects
at $M_P$.  The size of these effects depends upon
what fields carry the PQ--symmetry, but in general these
effects will give symmetry breaking corrections to
the axion (mass)${}^2$ of order $f^{p+2}/M_P^{p}$.  In order to avoid
upsetting the delicate cancellation of $\theta$ to
the precision of $10^{-9}$ we require
$f^{p+4}/M_P^{p} \lta 10^{-9}\Lambda^4_{QCD}$ or $p \gta 9$
using the parameters described above.  We would naively expect
$p \approx 2$ (or smaller!) in a scalar field effective theory.
Hence, the theory of the axion has a major hurdle to overcome
in being reconciled with the large ``$\theta$-pull''
induced by Planck scale effects.  

As we will
show in the present paper, the effective Lagrangian
of a latticized extra dimension offers a simple solution to this
problem. It also provides a natural origin
for the axion: the axion can be viewed as
the $5$th component of the
$U(1)_Y$ weak hypercharge gauge group in a world with a
periodically compactified
extra dimension.  It can alternatively
be viewed as arising in a natural
way, immune from Planck scale physics, within a
particular $1+3$ dimensional generalization of the Standard Model,
where $U(1)_Y$ is embedded into a chain
of $N$ gauge groups, $U(1)\times U(1)\times\cdots\times U(1)$
\cite{earlier,yanagida}.  A viable minimal scheme  
exists for $N$ as small as $N\sim 14$.

Ultra-low mass (pseudo)scalars are also
desired to solve a number
of cosmological problems, going
under the rubric of ``quintessence''  (for some recent
cosmo-phenomenological discussion, see e.g., \cite{silk}).
Peebles and Ratra \cite{Peebles:1988ek}
proposed a fundamental scalar theory,
where no symmetries are present,
which is one of the first ``quintessence''
models. Though a useful phenomenological construction,
from a fundamental point of view
this model suffers serious unnaturalness:
{\em (i)} the cosmological constant is tuned
to zero at the minimum of the potential;
{\em (ii)} the ultra-low mass scale is
put in by hand, and is otherwise
arbitrary and unmotivated; {\em (iii)}
Planck scale effects in such a theory would be expected
to pull the scalar mass term up to $\sim M_P$.
The cosmological constant problem {\em (i)} is common to all approaches
to issues in cosmology and particle physics, and we  offer no insight
into its resolution at present.

Independently, the other earliest  
``quintessence'' model, and a theory of late-time phase
transitions, was based upon established concepts of
spontaneous symmetry breaking
in particle physics. It  was proposed by
Hill, Schramm and Fry, and subsequently developed in
detail with Frieman, Gupta, Holman, Kolb, 
Stebbins, Waga, and Watkins,
\cite{ross,hillc,kolb,stein}.\footnote{In fact, the
idea of coupling these fields to $F_{\mu\nu}F^{\mu\nu}$
for electrodynamics and QCD was elaborated
in \cite{ross} and applied in \cite{stein} to constrain
time dependent fundamental constants. }
These models contain
an ultra--low mass PNGB, with a characteristic
potential of the form
 $\sim m_\nu^4\cos(\chi/f)$. Here $f$ ranges
from $f \sim M_{GUT}\sim 10^{15}$ GeV to $f\sim M_{P}$, the
scale of {\em spontaneous symmetry breaking}, 
and$m_\nu$ is a ``neutrino mass scale,'' or some
other comparable low mass scale, the scale of {\em explicit
symmetry breaking.} 
Expanding the potential
to quadratic order we see that
$\chi$ has the cosmologically interesting mass of $m_\nu^2/M_{P}$,
and a feeble coupling $\lambda\sim m_\nu^4/M^4_{P}$.
These small parameters were installed by hand in
the Peebles-Ratra model, but arise automatically here,
under control of symmetries, and
thus, the arbitrariness problem  {\em (ii)} described above does not  
arise in a PNGB scheme.

With regard to problem {\em (iii)} above,
the proposal of ultra-low mass PNGB's has
also been criticized (e.g., \cite{Carroll}), again
based upon  Planck-scale explicit symmetry breaking effects
in complete analogy to the axion.
Here the problem is more severe than in the case
of the axion because $f\sim M_{P}$ and there is evidently
no power law suppression to terms of the form $f^{p}/M_{P}^p$,
and the PNGB can apparently never have low mass.

In the present paper we
will show that
the Planck scale effects
for quintessence PNGB's are likewise
suppressed when they are reinterpreted
as $U(1)$ gauge fields periodically
compactified in an extra dimension. 
We will see that the bound on Planck scale effects 
typically restricts $f_\chi$ to
be less than $\sim$ a few $\times 10^{17}$ GeV.
Indeed,
in the minimal quintessence theory we present below,
we actually use the Planck scale effects to
generate the ultra-low mass of the PNGB.
This latter result appears to us to be very
natural within the framework of
the parameters of the deconstructed theory.\footnote{For 
another recent
proposal for quintessence fields arising from extra
dimensions see, e.g.,  \cite{burgess}.}

The key idea to solving the naturalness
problem of ultra-low-mass PNGB's
is to replace the PNGB by a gauge field in a
periodically compactified extra
dimension. Consider a
$U(1)$ gauge theory
in $1+4$ dimensions. This theory
contains the normal vector potential, $A_\mu$,
where $\mu = 0,1,2,3$,
plus a fifth component, $A_4$. We take
$x^4$ to be periodically compactified.
The $A_4$ gauge field, at low energies
(i.e., no momentum in the $x^4$ dimension), is
then effectively
a ``Wilson line'' which wraps around a compact
extra dimension, $\int_{loop} dx^4 A_4$.  This object
has the effective Lagrangian
of the PNGB of a global $U(1)$ symmetry.
We will refer to it as the Wilson Line PNGB (WLPNGB).

We can derive the effective
Lagrangian for the WLPNGB in $1+3$ dimensions
using the technique of latticizing (``deconstructing'') the extra
dimension \cite{HPW,wang0} (see also 
\cite{ACG}). The case of $U(1)$ electrodynamics
with a heavy fermion was
studied in detail in a previous
companion paper ref.~\cite{adam},
and will play a major role in the present discussion. 
The lattice provides a useful tool 
for regulating the enhanced quantum loop divergences of the 
extra dimensional theory, and generates a gauge 
invariant low 
energy effective Lagrangian with a finite subset  
of Kaluza-Klein modes.  A lattice description
of a $1+4$ theory involves chopping
the fifth dimension, $x^4$, into $N$ slices, or
``branes,'' each brane describing the $1+3$ spacetime
with its own copy of the gauge group.  $N$ plays the
role of a UV cutoff, and cannot exceed $\sim 4\pi/\tilde{\alpha}$,
where $\tilde{\alpha} = \tilde{g}^2/4\pi$ is the low energy effective
coupling constant of the $U(1)$ theory in $1+3$ dimensions
(the coupling ``runs'' to large values, $N\tilde{\alpha}=\alpha_{max}$
at short distances,
and $\alpha_{max}$ is typically
the unitarity bound of the coupling constant
\cite{HPW,chivukula}).

 The effective
Lagrangian in $1+3$  appears as $N$ copies of a
$U(1)$ gauge theory, of the form
$U(1)\times U(1) \times \cdots\times U(1) \equiv U(1)^N$.
The $U(1)$'s are linked together by $N$ Higgs fields,
$\Phi_n$, each having a common VEV.
When each $U(1)$ has the same coupling constant, and each
$\Phi_n$ the same VEV, there is a $Z_N$ symmetry
which replaces the continuous translational invariance
in the extra dimension. In fact,
one can simply forget about the
existence of an extra dimension and
view this as a procedure for constructing a new kind
of theory in $1+3$ dimensions, with a new symmetry group,  $Z_N$.
This procedure is conceptually powerful, and has suggested  
new directions and dynamics in building models 
of physics beyond the Standard Model, and provides a
new rationale for dynamical electroweak
symmetry breaking
(see, e.g. \cite{dob}). 
 
Planck scale effects are suppressed because
the WLPNGB is contained in
the $Z_N$ invariant product of the linking
Higgs fields,
$\widetilde{\Phi} = \Phi_1\Phi_2 \cdots\Phi_N$.
 Only the $Z_N$ invariant
field $\widetilde{\Phi}$
is gauge invariant.
Powers of any individual factor field, i.e., $\Phi_n^p$,
are gauge dependent, owing to the local $U(1)_n$,
and cannot be generated in the Lagrangian by
quantum gravity.  Symmetry breaking effects
from Planck scale physics must be gauge invariant
and can involve only
$\widetilde{\Phi}^p$, hence terms like:
\beq
\sim \frac{(\Phi_1\Phi_2 \cdots\Phi_N)^p}{M_P^{pN-4}} + h.c.
\eeq
These terms are seen to be
very highly suppressed owing to the dimensionality
of $\widetilde{\Phi}$. Such terms do produce
symmetry breaking effects (e.g., corrections to
the PNGB mass)
that are of order $(f/M_P)^{pN}$.  For large $N$
this provides the required suppression
of Planck scale breaking.

A conventional mass of the WLPNGB can also come
through its couplings
to matter in the bulk.  We have computed
the Coleman-Weinberg potential for the WLPNGB with
a heavy Dirac fermion propagating in the $1+4$
bulk,
and studied the anomalous couplings of the WLPNGB
to other gauge fields \cite{adam}.
The Coleman-Weinberg
effective potential for the WLPNGB
is finite when $N\geq 3$.
Finite potentials for PNGB's similar to these were discussed
long ago by Hill and Ross \cite{ross}. The 
finiteness is a consequence 
of the $Z_N$ invariance of the full theory. The 
``schizon'' models of Hill and Ross \cite{ross},
exploited $Z_{2L}\times Z_{2R}$
to reduce the degree of divergence from quadratic to
logarithmic and implement ultra-low-mass
PNGB's to provide natural ``$5$th'' forces
in the Standard Model, and remedy certain limits
on the axion.
With $Z_3$ symmetry,
finite neutrino-schizon models have been used 
to engineer the first ``quintessence''
models, late-time cosmological
phase transitions, and place limits upon time dependent
fundamental constants \cite{hillc}.  The
finite temperature behavior
of such models is also striking \cite{kolb}.
These models are structurally equivalent
to the present extra-dimensional scheme with
latticized fermions when written in the 
momentum space expansion in the fifth dimension \cite{adam}.

The present approach in principle solves a second  outstanding
issues for the axion: Why should the axion exist
at all?  In our present view
the axion is the $U(1)_Y$ gauge field
propagating in a compact extra dimension.  Indeed,
there does exist the $U(1)_Y$ in nature,
and the axion can occur if
there also exists the periodic extra dimension.
Probing the physics of the axion, e.g., measuring
its decay constant $f_a$, is a direct probe of
the physics of the extra dimension, e.g., $f_a \propto 1/R$ where
$R$ is the circumference of the extra dimension.

We take up the problem of constructing a bona-fide
theory of the axion as a consequence of the existence of
Standard Model $U(1)_Y$
and an extra compact dimension in Section 3.  We begin in Section
2 with the simpler problem of a minimal PNGB theory for
quintessence or late-time phase transitions.

 \vspace*{1.0cm}

\section{The Wilson Line Pseudo-Nambu-Goldstone Boson (WLPNGB)}  
 
\subsection{Gauge Field Lattice} 

Consider a free $U(1)$ gauge theory 
in $1+4$ dimensions that is {\em periodically 
compactified} to $1+3$:
\beq
\label{contin}
{\cal{S}}= \int d^4x\;\int_0^R\;dy\;\left[-\frac{1}{4} F_{AB} F^{AB} +
{\cal{L}}_P\right],
\eeq
where $(A,B)$ run from $0$ through $4$, and $y=x^4$.
${\cal{L}}_P$ describes the Planck scale symmetry
breaking effects, which we consider below.

The Wilson latticization  of the
gauge theory with a periodic fifth dimension is straightforward.
We slice the extra dimension into $N$ slices, or ``branes,''
each labeled by $n$.
The effective Lagrangian becomes  
the gauged chiral Lagrangian in $1+3$ dimensions for $N$ copies
of the $U(1)$ gauge group:
\be  
\label{lattone}
{\cal{S}}= \int d^4x \left[
-\sum_{n=1}^N\; \frac{1}{4} F_{n\mu\nu} F_n^{\mu\nu}
+ \sum_{n=1}^{N} D_{\mu}\Phi_n^\dagger D^{\mu}\Phi_n
+ {\cal{L}}_P \right].
\ee
In eq.~(\ref{lattone}) we have $N$ gauge groups,
$U(1)_n$, with  
a common gauge coupling  $g$,
and $N$ ``link-Higgs'' fields,
$\Phi_n$, having nonzero charges
in the $U(1)_n$ and $U(1)_{n+1}$
gauge groups only, as  $(0,0,...,1_n,-1_{n+1},0,0,...,0)$
(Note: $\Phi_N$ then has charges
$(-1,0,...,1)$, and we identify $n= N+1 = 1$, i.e.
$n$ is an integer mod$_N$).
Hence,  $\Phi_n$ ``links'' the $U(1)_n$ to the $U(1)_{n+1}$
gauge group. 
The covariant derivative therefore acts 
upon $\Phi_n$ as: 
\beq
D_{\mu}\Phi_n = \partial_{\mu}\Phi_n -
i g({A}_{{n+1}\mu}-{A}_{{n}\mu}) \Phi_n. 
\eeq
$g$ is the ``high energy'' value of the coupling
constant of the theory, while the low energy
coupling of the zero-mode photon is $\tilde{g}=g/\sqrt{N}$
\cite{HPW,wang0}, or $\tilde{\alpha} = \tilde{g}^2/4\pi
= \alpha/N$.

Each $\Phi_n$  has a vacuum expectation value, in which
the Higgs mode is very heavy, i.e., it
is effectively a nonlinear-$\sigma$ model field,
depending only upon its phase ${\chi}_n(x^\mu)$:
\be  
\label{phi}
\Phi_n \rightarrow (v/\sqrt{2}g)\exp(ig\chi_n / v). 
\ee 
The normalization is convenient and defines $v$
to be the inverse lattice spacing, $v=1/a$.
The VEV's
can easily be arranged with a judicious
choice of potential, common to each $\Phi_n$ \cite{HPW,wang0,adam}.

Through the equivalence of
eq.~(\ref{contin}) and eq.~(\ref{lattone})
we have mapped the $1+4$ theory into a
$1+3$ dimensional description. These two
theories are equivalent in their
low energy physics (modulo
certain lattice subtleties as addressed
in \cite{adam}).  Since we are only
interested in the extreme low energy limit,
this procedure is particularly useful, and
it buys a bonus: anything we say about the
low energy physics of the $1+4$
dimensional theory of eq.~(\ref{contin}), also
applies to the equivalent $1+3$ dimensional theory,
eq.~(\ref{lattone}).
We can therefore use geometrical intuition
from $1+4$ to understand the behavior of
the $1+3$ theory and {\em vice versa}.

The $\Phi_n$ kinetic terms then go over to a mass-matrix 
for the gauge fields: 
\be 
\label{massmatrix}
\sum_{n=1}^{N} D_{\mu}\Phi_n^\dagger D^{\mu}\Phi_n  \longrightarrow
\half v^2\sum_{n=1}^N \left| \left( ({A}_{n+1\mu}- {A}_{n\mu}) - 
 \frac{1}{v}\partial_\mu\chi_n  
\right)\right|^2
\ee  
Eq.~(\ref{massmatrix}) is easily diagonalized \cite{adam}.
Eq.~(\ref{lattone}) in the conjugate momentum space basis becomes:
\beq 
\int d^4x\left[\half (\partial_\mu\chi_0)^2 -
\sum_{p=-J}^{J} \; \frac{1}{4}F_{p\mu\nu}F^{p\mu\nu} +
 \sum_{p=-J}^{J} 2v^2 {A}_{p\mu}^2 \sin^2(\pi p/N) +{\cal{L}}_P \right],
\eeq 
(where without loss of generality consider
$N$ odd, and then $J = (N-1)/2$; see \cite{adam} for
more details).
The spectrum contains the zero-mass-mode $U(1)$ gauge field, $A_\mu$,
and the zero-mass-mode
$\chi_0$ (which is the zero-$p_4$-momentum $A_4$ or WLPNGB),
\begin{equation}
\chi_0 \equiv -\frac{iv}{g\sqrt{N}}  
  \ln \left[ \Pi_{n=1}^N(\sqrt2g\Phi_n/v) \right] =
\frac{1}{\sqrt{N}}\sum_{n=1}^N \chi_n.
\end{equation}
This mode is contained  in the nonlocal product of links.
It  corresponds to the continuum Wilson line around the 
compact 5th dimensions. 

The spectrum has  a tower of spin-$1$
Kaluza-Klein-modes,  appearing as massive gauge fields,
each labeled by $p$, of mass: 
\beq 
M_p^2 = 4v^2 \sin^2(\pi p /N). 
\eeq 
The tower of KK modes is linear in the large-$N$ limit.
All the $\chi_n$, except the zero-mode linear
combination $\chi_0$, are ``eaten'' 
to become longitudinal modes of the massive ``KK-photons''
\cite{wang0,adam}.

\subsection{Suppressing Planck Scale Effects }

We now include in the Lagrangian of eq.~(2.3,2.4) the
Planck scale breaking effects.
The leading allowed term is:
\beq
\label{grav}
{\cal{L}}_P =  
  \half\kappa \frac{e^{i\theta_P}}{M_P^{N-4}} \Pi_{n=1}^N \Phi_n  + h.c.
  + \dots
\eeq
where $\theta_P$ is an arbitrary ``gravitational CP-angle,''
an effective $\theta$-angle of the Planck scale vacuum.
$\kappa$ is a coefficient expected to be of order unity
associated with the details of Planckian physics.
This implies that the WLPNGB zero mode acquires a potential:
\beq
\label{grav2}
\kappa M_P^4 \left(\frac{v}{\sqrt{2}gM_P}\right)^N
\cos(g\sqrt{N}\chi_0/v + \theta_P)
\eeq
$g$ is the {\em high energy} value of the coupling constant
in this theory.

It is useful to recast eq.~(\ref{grav2}) in
terms of the low energy quantities in the theory.
We refer the reader to the companion paper \cite{adam}.
The low energy
value of the coupling, as would be
determined by the exchange of
low energy $U(1)$ gauge fields between currents
carrying any one of the $U(1)_n$ charges,
is \cite{wang0}:
\beq
\tilde{g} = g/\sqrt{N}; \qquad \tilde{\alpha} = \frac{\tilde{g}^2}{4\pi} = 
\frac{\alpha}{N}.
\eeq
From a continuum perspective, this is just
the power-law ``running'' of the gauge coupling in an
extra dimension. It arises in
the lattice description as a consequence of the mixing
of the chain of gauge fields in the $1+3$ effective Lagrangian.
The physical compactified circumference of the extra
dimension is given by:
\beq
R=N/v
\eeq
and the {\em decay constant} is \cite{adam}:
\beq
f_\chi = \frac{v}{g\sqrt{N}} = \frac{1}{\tilde{g} R}
= \frac{1}{\sqrt{4\pi \tilde{\alpha}} R}.
\eeq
The maximum value that $N$ can achieve is dictated
by $\tilde{\alpha}$ and the upper bound
of $\alpha\lta \alpha_{max}$, to which our
theory applies. This  implies:
\beq
N_{max} = \frac{\alpha_{max}}{\tilde\alpha}.
\eeq
The free theory has no  upper bound,
in principle, but any
couplings to matter will lead
a unitarity bound on $\alpha_{max}$.
If we take $\alpha_{max}$ close to its unitarity
upper bound, we have chosen a fine grained
lattice with brane interspacing $1/v_{max}$,
where the hopping parameter $v_{max} = N_{max}/R$,
corresponding to the shortest distance scales
at which we can apply the present field-theoretic description
of the system. We must, at shorter distances, go over
to the overarching theory, e.g., string theory. The unitarity
bound is dependent upon the matter content and details of the
theory, but a
reasonable limit corresponds to the breakdown
of perturbation theory, expected when $\alpha\sim 4\pi$.
Hence, we would expect:
\beq
N_{max} \lta \frac{4\pi}{\tilde\alpha}.
\eeq
Using these definitions we can rewrite
the gravitationally induced potential eq.~(\ref{grav2}):
\beq
\label{pot3}
V = \kappa M_P^4\exp\left[-\left(\frac{\alpha_{max}}{\tilde{\alpha}}\right)
\ln\left( {\sqrt{\frac{2\tilde{\alpha}}{\alpha_{max}}}\frac{M_P}{f_\chi}}
\right) \right]
\cos(\chi_0/f_\chi + \theta_P).
\eeq
We see that the induced potential is exponentially
suppressed for large $N$,  
provided we satisfy the requirement:
\beq
f^2_\chi < \left(\frac{2}{N}\right) M^2_P
=\left(\frac{2\tilde{\alpha}}{\alpha_{max}}\right) M^2_P.  
\eeq
If we use the largest possible value for $\alpha_{max} \approx 4 \pi$,  
we have the bound $f^2_\chi\lta (\tilde{\alpha}/2\pi)M_p^2$.  
Hence, as we would expect, there is
an upper bound on the decay constant of
a soft-PNGB such that Planck scale effects
are suitable suppressed.In fact the constraints are
nontrivial and arbitrary suppression is not possible.
This result is dependent upon the cut-off $N$ because the
leading term in  ${\cal{L}}_P$  is $N$ dependent,
and $\tilde{\Phi}$ is a nonlocal object.

A weakly coupled theory, $\tilde{\alpha} \ll 1$, can have large  
 $N_{max}$ and
negligible Planck scale corrections, provided $f_\chi$
is also taken small. As we will see below, this is the
case for the axion with its normal parameter
range, $f_a\sim 10^{12}$ GeV and with $\tilde{\alpha}\sim \alpha_{Y}$,
the coupling of the $U(1)_Y$ weak
hypercharge.
Taking
larger
values of $f_\chi$ requires larger
$N_{max}$ to enhance the $Z_N$ symmetry.

The mass of the WLPNGB, owing solely to Planck scale effects
is obtained by expanding the potential
to quadratic order about its minimum:
\beq
m_\chi^2 =
\kappa (M_P^4/f_\chi^2)
\exp\left[-\left(\frac{\alpha_{max}}{\tilde{\alpha}}\right)
\ln\left(
\sqrt{\frac{2\tilde{\alpha}}{\alpha_{max}}}\frac{M_P}{f_\chi}\right)
\right].
\eeq
While the result is expected in a true extra-dimensional theory
of the kind we have specified, it also applies
in $3+1$ theories having the structure of our
effective Lagrangian.
We can abandon the notion of the parent $1+4$ theory
and view this construction as a purely $1+3$ theory. This
is the philosophy of ``deconstruction''
advocated in \cite{ACG}.  In either case, $N$ is a parameter
which we are free to vary up to $N_{max}$.

\subsection{Minimal Theory of Quintessence}

A minimal model of quintessence can be constructed
by using a $1+4$ dimensional $U(1)$ pure gauge theory,
as in  eq.~(\ref{contin}), in a compact
extra dimension, or its equivalent
$1+3$ dimensional latticized description
eq.~(\ref{lattone}), as a source for the WLPNGB.
We couple the
$U(1)$ theory to a single Dirac fermion,
which also propagates in the bulk extra dimensions,
suitably latticized, below. For
the moment,
we neglect the effects of the fermion.
We tune
the coupling constant so that the ultra-low mass
of the WLPNGB is given by the Planck
scale effects.
As we see below,
the matter fields can produce additional
effects, but these are not problematic for the
present scheme.

The potential is therefore
given by
eq.~(\ref{pot3}).
Let us take $\kappa=1$.
If we saturate the unitarity bound of $\alpha_{max} \lta 4\pi$
we see that the potential becomes:
\beq
\label{pot4}
V = M_P^4 \exp\left[-\left(\frac{4\pi}{\tilde{\alpha}}\right)
\ln\left( \sqrt{\frac{\tilde{\alpha}}{2\pi}}
\frac{M_P}{f_\chi}\right) \right]
\cos(\chi_0/f_\chi + \theta_P).
\eeq
If this potential represents quintessence, it must
have an overall scale of order the closure density,
$V\sim\rho_c\sim 3H_0^2/8\pi G_N$,  
and we choose $H_0= 100 h$ (km s$^{-1}$ Mpc$^{-1}$) with $h=0.7$,  
hence $\rho_c = 4\times 10^{-47}$ GeV$^4$. 
Let $y=\ln(M_p/f_\chi)$ and
$x = 4\pi/\tilde{\alpha}$. Then:
\beq
y = \frac{1}{x}\ln(M_P^4/\rho_c) + \half \ln(x) - \half \ln(2).
\eeq
Thus the minimum value of $y$ is:
\beq
x = \frac{4\pi}{\tilde{\alpha}} = 2\ln(M_P^4/\rho_c),
\qquad
y = \ln(M_p/f_\chi) = \half\left(1+ \ln[\ln(M_P^4/\rho_c) ]\right).
\eeq
Note that $\ln(M_P^4/\rho_c) \approx 2.83\times 10^2$,
and we thus see that the optimal value
of $N_{max} \sim \alpha_{max}/\tilde{\alpha}  
\sim 4\pi/\tilde{\alpha} \sim 5.65\times 10^2$,
or $\tilde{\alpha} \approx 0.022$.
We can compare
$\tilde{\alpha}$ to the $U(1)_Y$ coupling constant of
the Standard Model, $\alpha_Y \sim 0.01$, and we see
that $\tilde{\alpha}$ is physically reasonable.
Moreover, we have
$f_\chi \sim 0.036\times M_P \sim $
or $f_\chi \sim 4 \times 10^{17}$ GeV.
This causes the quintessence field to have a
mass, $m_\chi \sim \sqrt{\rho_c}/f_\chi  
\sim 1.6\times 10^{-41}$ GeV,  
or,
expressed as a Compton wavelength,
of order $\sim(400\;{\rm Mpc})^{-1}$.
Larger values of the overall  
scale of the potential can be chosen, and
correspondingly smaller $f_\chi$ and larger
$\chi$ masses can be generated,  
since the field can relax while rolling in  
its potential.
  
We have provided only  
a crude estimate, and we will not presently give a
detailed ``cosmic evolution'' of the $\chi$ field.  
We remark,
however, that the scale of the potential is so sensitive
to the values of these parameters that
it is conservative to say that
$f_\chi \gta \makebox{few} \times 10^{17}$ GeV
would yield an unacceptably large vacuum energy in
the $\chi$ field today.
For these
parameters we have a field that may begin rolling
in its potential at redshifts $z \sim 10$.
This is roughly in accord with
the Late Time Phase Transition
of Hill, Schramm and Fry, \cite{hillc}.
A network of soft domain walls may form.
The extent to which $\chi$ can be trapped in its
potential, e.g., dwell near the maximum
until redshifting like matter, thus providing
an effective cosmological constant, remains to
be explored. The cosmic evolution of a quintessence
field $\chi$ with these parameters is subtle
and beyond the scope of the present paper.

One thing we learn immediately from this
exercise is that it is evidently not possible
to have a PNGB field that {\em naively}  provides a
cosmological constant today by being
trapped in its potential due to Hubble damping.
To achieve this condition
we would require a potential of the form:
\beq
\sim \Lambda_c\cos(\chi/f)
\eeq
and the mass would have to satisfy
$m_\chi^2 =\Lambda_c/f^2\lta H^2 \sim \Lambda_c/M_p^2$,
hence we would require $f\sim M_P$.  However,
this requires $\tilde{\alpha}\rightarrow\alpha_{max}$.
and we cannot adequately suppress the Planck effects
in this case. We suspect that this objection
may be more general, and apply to any
natural quintessence model.  It also implies
difficulties for PNGB
inflaton models \cite{freese},
which typically require $f\sim M_P$ and Hubble
damping to provide a cosmological constant
phase.

The resulting overall scale of the potential
for this ``pure quintessence'' model
is extremely sensitive to the parameters we
have used, and there is thus considerable fine-tuning
of such a model.  It would evidently be better to raise
$\alpha_{max}/\tilde{\alpha}$, or $M_P/f_\chi$
slightly and thus turn the effects of Planck scale
physics off completely. In this case we require some
other mechanism to generate the small
WLPNGB mass. This can come from fermions
to which the $U(1)$ theory couples.
Examples are the neutrino schizon models
described in refs.~\cite{hillc,kolb}, though
some additional $U(1)^N$
and $Z_N$ apparatus must be added
to satisfy the constraints described here.
We examine presently an alternative
scheme, a fermion which
``propagates in the bulk.''

\subsection{Effect of Fermions in the Bulk}

Let us now incorporate a Dirac fermion which
can be viewed as carrying the $U(1)$
charge and propagating in the bulk. 
The continuum
action is:
\beq
\label{contferm}
{\cal{L}} = \int d^4x \; \int_0^R\;dy\;
\bar{\Psi}[(i\slash{\partial}-g\slash{A})
-(\partial_4+igA_4)\gamma^5 - M ]\Psi.
\eeq
Latticizing
and passing to a chiral  basis  of $L$-handed
and $R$-handed fields
on each brane, the theory can be written
({\em c.f.} eq.~(2.37) of \cite{adam}):
\bea
& & \sum_{n=1}^N\int d^4x \;\left[  \bar{\Psi}_{nL}(i\slash{D}
 )\Psi_{nL}  
 + \bar{\Psi}_{nR}(i\slash{D}
) \Psi_{nR} - \widetilde{M} (\bar{ \Psi}_{nL} \Psi_{nR} + h.c.) 
\right] \nonumber \\  
 & &
-{\sqrt{2}}  \sum_{n=1}^N\int d^4x \; g 
\left[\eta_1\bar{ \Psi}_{nL} \Phi_n \Psi_{(n+1)R} -
\eta_2\bar{ \Psi}_{nR} \Phi_n  \Psi_{(n+1)L} + h.c.\right]
 \label{latferm1}
\eea
This lattice action
for the Dirac fermion is discussed in detail in \cite{adam}.
Here $\Psi_{R,L} = \half(1\pm\gamma^5)\Psi$
and $\widetilde{M} = M-\eta v$, where $M$
is the continuum theory's Dirac mass
of eq.~(\ref{contferm}).
eq.~(\ref{latferm1})
represents a faithful lattice description of the continuum
theory. We have included the ``Wilson term'' which eliminates
an unwanted spurious fermion flavor doubling
in the spectrum. 
When the latticized model
is constrained to match the low
energy spectrum of the continuum theory,
we find a nontrivial constraints on $\eta_i$ \cite{adam},
\beq
\eta_2 \ll \eta_1\equiv \eta.
\eeq
Note that we can swap
$\eta_2\leftrightarrow\eta_1$ under a parity
transformation.
The matching of the spectrum further requires:
\beq
v^2 = -\eta v \widetilde{M} = \eta v(\eta v - M),
\qquad \eta = \frac{M\pm\sqrt{M^2 + 4v^2}}{2v}.
\label{result10}
\eeq
If we implement these constraints, and
substitute $\Phi_n = (v/\sqrt{2} g)\exp(ig\chi_n/v)$,
the fermionic
lattice action can be rewritten in terms
of  the parameters $v$ and $\eta$, as:
\bea
& & \sum_{n=1}^N\int d^4x \;\left[  \bar{\Psi}_{n}(i\slash{D}
 )\Psi_{n}+ \left(v/\eta\right)\bar{\Psi}_{n} \Psi_{n}
-(\eta v\bar{ \Psi}_{nL}\Psi_{(n+1)R}e^{ig\chi_n/v} 
+ h.c.)\right]
 \label{latferm2}
\eea 
We can regard eq.~(\ref{latferm2}) as a $1+3$
dimensional model with effectively two mass
parameters, $m_1 = v/\eta$ and $m_2 = \eta v$.
A third allowed set of terms,
$\eta_2 v \Psi_{nR}\Psi_{(n+1)L}e^{ig\chi_n/v}$,
can in principle occur, but we assume
$\eta_2 \ll \eta$ by at least an order of magnitude,
and does not affect the present estimates.

The fermion couplings to the Higgs-link fields
of eq.~(\ref{latferm1}) cause the
WLPNGB to develop a Coleman-Weinberg effective
potential \cite{adam}.
For $\eta \approx \pm 1$
the $\chi_0$ acquires
a large mass of order $\tilde{\alpha}v^2/N^2 \sim
\tilde{\alpha}^2 f^2_\chi\sim \tilde{\alpha}/R^2$.
In this limit the WLPNGB cannot be an ultra-low-mass
spin-$0$ field if
the compactification scales, $R^{-1}= v/N$, are at least as
large as $\sim$ many (TeV)$^{-1}$.

For
large or small $\eta$ we see that either the
chiral breaking $m_1$ terms disappear, or
the chiral coupling $m_2$ terms disappear,
and the Coleman Weinberg effective potential
becomes small. It actually does not matter which limit to use,
since the Coleman-Weinberg potential depends
symmetrically upon $\eta^2 + \eta^{-2}$ and $v$ in
this limit.
Define:
\beq
\omega = \sqrt{\eta^2 + \eta^{-2}}.
\eeq
In the large $\omega$
limit we obtain
an exponentially suppressed
Coleman-Weinberg potential \cite{adam}:
\beq
V = -\frac{v^4}{4\pi^2}\frac{\omega^{-2(N-2)}}{N^2}
\cos  
\left(g\sqrt{N}\chi_0/v\right)
\eeq
which can be written in terms of low energy parameters:
\beq
\label{quint}
V
= -\frac{\omega^4\alpha_{max}^2 e^{-(\alpha_{max}/\tilde\alpha)
\ln(\omega^2)}}{4\pi^2 R^4\tilde{\alpha}^2}
\cos  
\left(\chi_0/f_\chi\right)
\sim 
-4\pi^2 \left({\omega^4f_\chi^4}\right) e^{-(4\pi/\tilde\alpha)
\ln(\omega^2)}
\cos  
\left(\chi_0/f_\chi\right).
\eeq
In the latter
expression we used the
``unitarity bound,'' $\alpha_{max} =4\pi$.

We can thus choose the Planck-scale effects
to be arbitrarily small, and generate
the WLPNGB mass for quintessence from eq.~(\ref{quint}).
For example, with the choice of parameters
$x = 4\pi/\tilde{\alpha} = 10^2$
($\tilde{\alpha}\approx0.12$) and
$f_\chi \lta 6.3\times 10^{16}$ GeV, we see that
the Planck scale effects become miniscule.
We can then have the fermion contribution to the
potential dominate and provide vacuum
energy of order $\rho_c$ with
$\omega \sim 3.9$. This field has a mass  
of $m_\chi \sim 5.0\times 10^{-41}$ GeV, hence a  
Compton wavelength of order $\sim 120$ Mpc.

\section{Theory of the Axion}

The previous discussion shows that it is
possible to reduce Planck scale effects
well below the order of the closure density
with $f\lta 10^{-2}\times M_P$ and $\tilde{\alpha}$
of order the $U(1)_Y$ electroweak coupling constant.
This is a far better control than is required
to adapt the scheme to a theory of the axion.

Consider a scheme in which  we incorporate the
Standard Model gauge groups
$SU(3)\times SU(2)_L$ and all
matter fields and Higgs fields
into the latticized effective theory.
We place all of these Standard Model fields on a
particular lattice brane, e.g.,  $n=1$ without
loss of generality.
We allow, however, the $U(1)_Y$
to fill the bulk, i.e., we replace
$U(1)_Y \rightarrow [U(1)_Y]^N$, hence we
take the $U(1)_Y$ Lagrangian to be of
the form of  eq.~(\ref{lattone}), describing a
latticized, periodically compactified,
extra dimension. One
can simply view this as a chain model of $U(1)$'s
in $1+3$ dimensions.

Hence, the model we are considering
from the $1+3$ perspective is
 $SU(3)\times SU(2)_L\times U(1)_Y^N$.
The weak hypercharges of the $U(1)_{Yn}$
will be denoted $Y_n$.  All Standard Model fermions and the
Higgs field couple only to $U(1)_{Y1}$ with the usual
weak hypercharges, $Y_q= Y_{q1}$. The $U(1)_{Y1}$
coupling constant is $g$, the common high energy
value of all $U(1)_{Yn}$'s, but the zero mode
gauge field will couple with strength $g_1=g/\sqrt{N}$,
hence:
\beq
 \alpha(M_Z)_{QED}^{-1} \sim 128;
 \qquad \tilde{\alpha}(M_Z) = \frac{\alpha(M_Z)_{QED}}{\cos^2\theta_W}
 \approx 0.01.
\eeq
This model reproduces as a low energy effective
Lagrangian, the Standard Model, with the zero-mode
gauge field of the $U(1)_{Yn}$ playing the role
of the usual $U(1)_Y$ gauge field
(for details of this kind of structure
see,  \cite{wang0}).

One might object to a periodic compactification of the
$U(1)_Y$ given the chiral fermionic zero-modes on brane-1.
Chiral fermions are usually engineered in extra-dimensional theories
with domain walls, and/or require ``orbifold'' compactification.
Orbifold compactification, unlike periodic
compactification, does
not produce a zero-mode $\chi_0$. The orbifold can, however,
be considered as a dynamical system. It can
be viewed as a ``parallel brane capacitor''
bounding an extra dimension, with the
branes at, e.g., $x^4 =0$ and $x^4=R$,
or a single brane with $x^4=0$ and $x^4=R$
identified. The capacitor branes
are a {\em Type II magnetic superconductor}, which is equivalent
to a {\em confining
phase} (magnetic superconductors admit
electric flux tubes which confine electric charge).  The gauge field
boundary conditions are accordingly $F_{\mu 4}=0$ at $x^4=0$ and
$x^4=R$.  This implies no electric (magnetic) field perpendicular
(transverse) to the brane at the boundary.
It also implies (i) $\partial_4 A_\mu=0$ and (ii) $\partial_\mu A_4 =0$
at the boundary.  (i) implies that there can exist a zero $p^4$
momentum mode of $A_\mu$ while (ii) implies that there cannot
exist a zero-$p^4$ momentum mode of $A_4$. Hence, $\chi_0=0$
with orbifold compactification.
From such a ``dynamical orbifold'' point of
view, certain fields, e.g.,
$SU(3)$, might propagate in the bulk as if an orbifold, and ``feel the
capacitor branes,'' while other fields such as the $U(1)_Y$ are decoupled
from the parallel branes and are periodic.
Hence, a
geometrical extra-dimension interpretation may
exist for our present model with dynamical orbifolding,
but we will not elaborate this presently.

Moreover, from a $1+3$, or ``theory space,'' point of view there
is no problem with such a construction, since it may be
viewed strictly as a $1+3$ theory with $N$ copies
of the $U(1)_Y$, and we summarize a viable minimal
scheme with $N=14$ below.
We thus use, for the gauge part of our $U(1)_Y^N$
theory, the Lagrangian of eq.~(\ref{lattone}).
The zero-mode of the $A_4$ gauge field in the continuum
version of the theory
is again the chiral phase in the product of linking
Higgs fields, $\Pi_{n=1}^N\Phi_n$
of the $[U(1)_Y]^N$ theory.
This phase, $\chi_0$, will be identified with the
axion. If no fermions propagate in the bulk,
the $\chi_0$ is massless, up to Planck scale effects.
The Planck scale effects are especially small,
much smaller than those considered
for quintessence,  given that
we typically want $f_\chi \sim 10^{12}$ GeV, and the $\ln(M_P/f_\chi)$
is becoming $\sim{\cal{O}}(10^1)$, large.
The problem, however, is that
the axion, without fermions, is decoupled from the
QCD field strength combination $G{}^\star{G}$,
where ${}^\star G_{\mu\nu} =
\half\epsilon_{\mu\nu\rho\sigma}G^{\rho\sigma}$.
This is required to provide the mechanism for
canceling the $\theta$ term.

Setting aside the problem of controlling
Planck scale effects momentarily,
we note that 
a particularly simple ``bare-bones'' theory of the axion could
always  be engineered
by postulating a heavy, e.g., $M\gg$ TeV, vector-like color
triplet whose mass is chiral,
$\sim M\bar{Q}_LQ_Re^{i\chi_0/f_\chi}+ h.c.$. This theory
is classically $U(1)_L\times U(1)_R$
invariant, but the $U_{R-L}(1)$ PQ-symmetry is broken
by the QCD axial vector anomaly. Rephasing $Q$
as $Q_L \rightarrow e^{i\chi/2f}Q_L$ and
$Q_R \rightarrow e^{-i\chi/2f}Q_R$ removes the
axion phase from the
mass term, and gives it a derivative coupling
to the axial current, but induces the Wess-Zumino term:
\beq
\label{anom1}
\frac{\alpha_{QCD}}{4\pi} \frac{\chi_0}{f_\chi} G^a_{\mu\nu}\;{}^\star
G^{a\mu\nu}.
\eeq
The point
is that it is unnecessary to directly couple the axion
into the light quark mass matrix.  Through the effects
of the anomaly in QCD, the W-Z term
of eq.~(\ref{anom1}) will lead
to mixing of the $\chi$ field with the $\eta',\eta,\pi^0$
and a mass for $\chi$ will be induced of order
$\Lambda_{QCD}^2/f$ (this mixing also induces
the $(\chi_0/f_\chi) F{}^\star F$ anomalous electromagnetic
coupling). The VEV of $\chi_0$ at the potential
minimum will identically cancel the
unwanted QCD $\theta$ angle.

It is actually not necessary that the PQ-symmetry
be classically exact. One can add
(before rephasing) a ``small chiral breaking'' term
of the form $\delta m \bar{Q}_LQ_R + h.c.$.  This would
induce in loops a quadratically divergent axion mass,
$\sim \Lambda^2\delta m M \cos(\chi/f)$.
However, we know
how to soften this by invoking a $Z_N$ symmetry, with more
$Q$ fields, so we can ignore it presently.
This, however, affects the anomaly, since it
redefines $\chi$ as:
\beq
\overline{M} e^{i\chi'/f} = M e^{i\chi/f} + \delta m,
\qquad \overline{M}^2 = M^2 +(\delta m)^2 + 2\delta m M \cos(\chi/f),
\eeq
and it is $\chi'$ that is ultimately
rephased and couples as in eq.~(\ref{anom1}).
This is a correction to the anomaly coming
from the explicit chiral breaking embodied in $\delta m$.
Note that as $\delta m/M \rightarrow 0$,
we have $\chi' \rightarrow \chi$.  As long as $\delta m$
is not too large, and the induced potential for
the axion is sufficiently negligible, this has
no effect  on the low energy dynamics.

For us, the simplest method
to engineer the necessary coupling
of the axion to $G^\star G$ is to imitate
this bare-bones model and introduce $N$ copies of
a vector-like (heavy) quark color triplet $\Psi_{n}$ with
sequentially different weak hypercharges,
$(0,0,...0,-1_n,0,0..0)$ for $(U(1)_{Y1},...,U(1)_{YN})$.
One can view this as a fermion
that propagates in the bulk. We could also allow $SU(3)$
to propagate in the bulk,
however, it is convenient to simplify the model, and
simply assign each of these fields to be a color triplet
relative to single $SU(3)_{QCD}$ gauge group.
We thus have an anomaly free representation of $\Psi_n$'s
under the full  $SU(3)_{QCD}\times SU(2)_L\times U(1)_Y^N$.

This common color triplet
assignment indeed appears to have nothing to
do with a geometrical theory.  To make contact
with geometry, we require QCD itself
propagate in the bulk, and to have the lattice
description $[SU(3)]^N$. 
Indeed, we have an impetus to allow QCD in the bulk: 
Topcolor, and in particular, the Top Seesaw model \cite{dob},
is a successful dynamical scheme for
breaking the weak interactions in analogy
to BCS theory, and can be viewed
as an extra dimensional theory with $SU(3)$
and the top quark propagating in the bulk.
One can view the present common triplet model as a
technical simplification
of a more complete QCD-in-the-bulk theory.

The common triplet scheme is sufficient to generate
the required anomalous coupling of the axion. Again
we have the Lagrangian of eq.~(\ref{latferm1},\ref{latferm2}).
The couplings to the axion,
which is the $\chi_0$ zero mode, are:
\bea
& & \sum_{n=1}^N\int d^4x \;\left[  \bar{\Psi}_{n}(i\slash{D}
 )\Psi_{n}+ \left(v/\eta\right)\bar{\Psi}_{n} \Psi_{n}
 \right]
 \nonumber \\
 & & +\sum_{n=1}^N\int d^4x \left[  
(-\eta v\bar{ \Psi}_{nL}\Psi_{(n+1)R}
+ \eta_2 v \bar{ \Psi}_{nR}\Psi_{(n+1)L}
)e^{i\chi_0/f_\chi} + h.c.\right]
 \label{latferm3}
\eea
Again we assume $\eta_2\ll\eta_1 \equiv \eta$,
the form apropos the Wilson term, but we emphasize that
$\eta_2$ need not be hierarchically smaller than $\eta_1$.

The $(v/\eta)$ diagonal terms which explicitly break the
$U(1)_{PQ}$ chiral symmetry
are the analogue of $\delta m$ described
above. We would require fixed $\eta v$
and $v/\eta \rightarrow 0$
to recover exact chiral symmetry, whence the
Coleman-Weinberg potential would vanish.

The Coleman-Weinberg potential is
given as before
in terms of $\omega = \sqrt{\eta^2 + \eta^{-2}}$, for
$\omega$ large,
by the exponentially suppressed potential:
\beq
V
\sim 
-4\pi^2 \left( {\omega^4 }{f_\chi^4}\right) e^{-(4\pi/\tilde\alpha)
\ln(\omega^2)}
\cos  
\left(\chi_0/f_\chi\right).
\eeq
where we have used $\alpha_{max}=4\pi$.
  
The dominant contribution to the axion potential  
will come from QCD, due to the usual instanton  
effects.
The key to our present theory
is that we must insure the axion is not pulled away
from its QCD potential minimum,
where $\theta_{QCD}$ is canceled, by more than $1:10^{-9}$.
We thus require:
\beq
\label{cond1}
4\pi^2 (f_\chi^4\omega^4) e^{-(4\pi/\tilde\alpha)\ln(\omega^2)}
\lta 10^{-9}\Lambda^4_{QCD}.
\eeq
Using  $f_\chi\sim 10^{12}$ GeV,
and $\tilde{\alpha}\sim {\alpha}_Y \sim 0.01$, we see that
$\omega\gta 1.1$ causes the {\em lhs}
of eq.~(\ref{cond1}) to  become negligible.  Since 
$\omega\ge\sqrt{2}$, any value of $\eta$ may be used,  
i.e., $\eta$ need not be tuned excessively large
or small. More
relevant is the validity of our large-$\omega$ approximation;
we expect that it requires $\eta \sim \omega \gta 10$.
Moreover,
the Planck scale effects must also  
be suppressed to a comparable level:
\beq
M_P^4e^{-(4\pi/\tilde{\alpha})
\ln[ \sqrt{\tilde{\alpha}}M_P/\sqrt{2\pi}f_\chi)] }
  \lta 10^{-9}\Lambda^4_{QCD}.  
\eeq
With the parameters $f_\chi\sim 10^{12}$ GeV,
and $\tilde{\alpha}\sim {\alpha}_Y \sim 0.01$ we see that
the {\em lhs} is negligible.

Since the Planck scale effects are miniscule
with $f_\chi =10^{12}$ GeV,  
we might ask
what is the smallest value of $N = \alpha_{max}/\tilde{\alpha}$
we can choose
such that both the Planck scale and $\Psi_n$
$\theta$-pull effects are
still negligible?
Our underlying theory will contain a  
fermion mass term $v\eta$ or $v\eta^{-1}$  
in excess of $M_P$ unless  
we also enforce 
$\omega \lta M_P/v \sim M_P/\tilde{g}{N}
f_\chi\sim 3.4\times 10^7/{N}$.
The Planck scale $\theta$-pull  
effects are:
\beq
M_P^4e^{-N
\ln[\sqrt{2}M_P/\sqrt{N}f_\chi] }
 \lta  10^{-9}\Lambda^4_{QCD}  
\eeq
We see that $N>13$ suffices to reduce  
these to a negligible status with $f_\chi\sim 10^{12}$ GeV.
Then the fermion $\theta$-pull effects require:
\beq
\label{cond2}
4\pi^2 (f_\chi^4\omega^4) e^{-2N\ln(\omega)}
 \lta 10^{-9}\Lambda^4_{QCD}
\eeq
Hence, with $N= 14$  
the $\theta$-pull effects are satisfied with the range  
$4.1\times 10^2\lta\omega\lta 2.5\times 10^6$. Thus, the minimal  
model has $N=14$ copies  
of $U(1)_Y$, with the  
low energy coupling $\tilde{\alpha}=\alpha_Y=0.01$,  
$f_\chi = 10^{12}$ GeV.  Using $\omega = 4.1\times 10^2$,
the Lagrangian of eq.~(\ref{latferm3})  
has the chirally invariant mass term $v\eta \approx 2\times 10^{15}$  
GeV and the chiral breaking term $v/\eta \approx 10^{10}$ GeV.

Computing the loops involving the
$\Psi_n$ generates an anomalous coupling
of $\chi_0$ to two gluons, and to pairs
of the $U(1)_Y$ field and its KK-modes \cite{adam}. 
The anomalous coupling
of $\tilde\chi_0$ to the $[U(1)_Y]^N$ was computed
in the companion paper \cite{adam}, and the QCD
coupling is similarly easy to derive.
We obtain:
\beq
\frac{\eta\chi_0}{16\pi^2 \omega f_\chi} \left[
4\pi N \tilde{\alpha}_3 G^a_{\mu\nu}{}^\star {G}^{a\mu\nu}
+ 4\pi\tilde{\alpha}_1\sum_{n=1}^N F_{n\mu\nu}{}^\star {F}^{a\mu\nu}
\right].
\eeq
Notice the large coefficient of $N$ in the
QCD piece. This coefficient would be unity if
we allowed QCD to propagate in the bulk, and
$G^a_{\mu\nu}{}^\star {G}^{a\mu\nu}
\rightarrow \sum_{n=1}^N G^a_{n\mu\nu}{}^\star {G}_n^{a\mu\nu}$.
The $\eta/\omega$ coefficient reflects the explicit
chiral symmetry breaking for small $\eta$, and the decoupling
of $\chi_0$ from the fermions. For large $\eta$, we
have $\eta/\omega \rightarrow 1$ and
the coupling is given purely by the anomaly.

In summary, from a $1+3$ dimensional (``theory
space'') point of view we have  built a natural
model of the axion based upon the
gauge group $SU(3)\times SU(2)_L\times U(1)_Y^N$
where, e.g., we can minimally
choose $N=14$. $U(1)_Y^N $ is broken by $N$
linking Higgs-fields $\Phi_n$ as in eq.~(\ref{lattone}).
The model includes $N$ vector-like color triplets
that carry sequential $U(1)_n$ charges
with the Lagrangian of eq.~(\ref{latferm3}).
The axion is then the chiral phase (zero-mode)
contained in the product $\tilde{\Phi} \propto \Pi_{n=1}^N \Phi_n$.
This
suppresses the Planck scale corrections to the axion
mass adequately to permit the cancellation of
the QCD $\theta$-angle to well within $1:10^9$, and to yield the
usual QCD-induced axion mass $\sim\Lambda_{QCD}^2/f_\chi$.
From an extra-dimensional point of view
the axion is the Wilson line wrapping around
a fifth compact dimension. More generally,
$N$ can be taken as large
as $4\pi/\alpha_Y \sim 1.3\times 10^3$ in
the latticized extra dimension scheme,  
or as small as $N\approx 14$.

\newpage
\section{Conclusions}

We have  generated naturally low-mass
pseudo-Nambu-Goldstone bosons, relatively
immune to the effects of Planck scale breaking
of global symmetries, and which can be
used for a number of phenomenological purposes.  These low mass
PNGB's are the Wilson lines wrapping around extra compact
dimensions in the presence of heavy fermions
that propagate in the bulk. Or, they
are products of  fields occurring
in an equivalent $1+3$ theory.
Because they are nonlocal
objects, Planck scale effects that
normally destroy global symmetries
are suppressed.

It is, in fact, somewhat remarkable that we can make
reasonable estimates of Planck scale effects in
these models and obtain nontrivial constraints.
This stems from using the lattice
approach and latticizing the physics of a
compact extra dimension, to rewrite it
in terms of a $1+3$ dimensional theory.
In a sense, latticization (or deconstruction)
which is normally viewed as a configuration space
cut-off,  is an
operator product expansion which isolates the
relevant components of the theory necessary
to describe its low energy physics.
The operator product expansion, which
reliably accommodates the scaling behavior of
various operators, is what allows us to
estimate the magnitude of the Planck scale effect
in the $1+3$ effective theory.

These fields can be quintessence fields.
We have not given a detailed analysis of
the cosmic evolution of these fields, but
they can yield interesting effects at late times
in the early Universe. Such fields have
been anticipated in earlier works \cite{hillc,kolb}
but until now the deleterious Planck scale effects
have not been addressed \cite{Carroll}.
  
We have also been able to solve the nagging problem
of Planck scale effects and the axion. With $f_a\sim 10^{12}$ GeV,
we can construct
models in ``theory space'' with $14\lta N\lta 10^3 $ heavy fermions and
an approximate $U(1)$ PQ-symmetry, sufficient to
yield an axion with no significant Planck effects,
with no significant $\theta$-pull,
thus allowing the axion to  cancel
the unwanted $\theta$-term of QCD.

A point worth emphasizing is that the Coleman-Weinberg
potential arises naturally in these schemes as a finite quantity
\cite{ross}.
There is no renormalization of the potential that is
implemented ``by hand,'' as in the case of massless
scalar electrodynamics. This suggests new directions in
thinking about improved models of the inflaton.

\section*{Acknowledgements}

We wish to thank S. Carroll for useful discussions
which largely stimulated the present work.
Research by CTH and AKL was supported by the U.S.~Department of Energy
Grant DE-AC02-76CH03000.

\vspace*{1.0cm}
  
\frenchspacing

\end{document}